\documentclass[ACS,STIX1COL]{WileyNJD-v2}

\articletype{Original Article}%

\received{}
\revised{}
\accepted{}

\newcommand{\kB}{k_\text{B}}

\newcommand{\wpl}{\omega_\text{p}}
\newcommand{\wc}{\omega_\text{c}}
\newcommand{\rL}{r_\text{L}}
\newcommand{\vth}{v_\text{th}}
\newcommand{\wuh}{\omega_\text{UH}}

\begin{document}
\title{Dynamic structure factor and excitation spectrum of the one-component plasma: the case of weak to moderate magnetization}

\author[1]{Hanno Kählert*}
\author[1]{Michael Bonitz}

\address[1]{\orgdiv{Institut für Theoretische Physik und Astrophysik}, \orgname{Christian-Albrechts-Universität zu Kiel}, \orgaddress{\country{Germany}}}

\corres{*Hanno Kählert, Institut für Theoretische Physik und Astrophysik, Christian-Albrechts-Universität zu Kiel, Germany. \email{kaehlert@theo-physik.uni-kiel.de}}


\abstract[Summary]{Magnetized plasmas are well known to exhibit a rich spectrum of collective modes. Here, we focus on the density modes in dense or cold plasmas, where strong coupling effects alter the mode spectrum known from traditional weakly coupled plasmas. In particular, we study the dynamic structure factor (DSF) of the magnetized one-component plasma with molecular dynamics simulations. Extending our previous results [H.~Kählert and M.~Bonitz, Phys. Rev. Research \textbf{2022}, 4, 013197], it is shown that Bernstein modes can be observed in the weakly magnetized regime, where they are found below the upper hybrid frequency, provided the coupling strength is sufficiently low. We investigate the DSF for a variety of different wave numbers and plasma parameters and show that even small magnetization can give rise to a strong zero-frequency mode perpendicular to the magnetic field and change the dispersion as well as the damping of the upper hybrid mode.}


\maketitle


\section{Introduction}\label{sec:intro}
Strong coupling effects can be observed in a variety of different plasmas, ranging from dusty plasmas~\cite{bonitz-introcomplex} to confined charges~\cite{dubinreview,danielson2015rmp,fajans2020pop} and expanding ultra cold neutral plasmas~\cite{killian1999,killian2012pop,lyon2016ultracold}. Even though these systems have very different density and temperature, they can all be characterized as strongly coupled, meaning that the Coulomb coupling parameter,
\begin{equation}
    \Gamma=\frac{Q^2}{4\pi \epsilon_0\,a\, \kB T},
\end{equation}
is close to or exceeds a value of one. The coupling is determined by the plasma density $n$ via the Wigner-Seitz radius, $a=[3/(4\pi n)]^{1/3}$, the temperature $T$, and the particle charge $Q$, showing that a large $\Gamma$ can be achieved with different combinations of $n$, $T$, and $Q$. Plasmas with the same $\Gamma$ can behave very similarly even when their physical parameters are vastly different.

While the properties of (classical) strongly coupled plasmas have been studied in great detail, mainly through simulations or theory of reduced model systems such as the one-component plasma (OCP), e.g., Refs.~\cite{hansen1973,hansen1975, tanaka1987pra, donko2000, korolov2015cpp, ichimarureview}, less attention has been paid to the characteristics of magnetized strongly coupled plasmas. The dense Coulomb liquids and solids in the crust of neutron stars~\cite{Potekhin2015ssr} or laser-cooled ions in Penning traps~\cite{anderegg2009prl,anderegg2010pop} can easily reach a regime with very high magnetization. In addition, ultra cold neutral plasmas have recently been studied under the influence of external magnetic fields~\cite{zhang2008prlmag,gorman2021prl,sprenkle2022pre,gorman2022pra}.  The strongly coupled dust particles in dusty plasmas are notoriously difficult to magnetize. However, setting the dusty plasma into rotation, such that the Coriolis force becomes appreciable, allows one to create an intense \textit{effective} magnetization~\cite{kaehlert2012prl,bonitz2013}, which has been utilized to study waves~\cite{hartmann2013} and transport phenomena in two-dimensional dusty plasmas~\cite{hartmann2019pre}. External magnetic fields also play an important role in the context of inertial fusion concepts~\cite{schmit2014prl,gomez2014prl,gomez2020prl,bose2022prl}. Previous \textit{theoretical} work for strongly coupled magnetized plasmas has been performed, e.g., for transport properties~\cite{ott2011prl,feng2014pre,ott2017pre,baalrud2017pre, vidal2021pop}, stopping power and friction~\cite{bernstein2020pre,jose2020pop,bernstein2021pop,jose2021pop}, and waves~\cite{hou2009pop,baiko2009pre, baiko2010jpcs, bonitz2010prl,yang2012pop,ott2012,kaehlert2013,ott2013}.

In this work, we focus on density correlations in the time and spatial domain by studying the dynamic structure factor (DSF) of the magnetized OCP. Specifically, we extend the results of Ref.~\cite{kaehlert2022prr} by mainly considering the DSF for weakly magnetized systems. The results should be particularly useful for experiments in which magnetization is difficult to achieve. This is the case, i.a., in very dense plasmas, where the high plasma frequency, $\wpl=\sqrt{n\,Q^2/(\epsilon_0 m)}$ (mass $m$), requires very intense magnetic fields to increase the magnetization parameter $\beta=\wc/\wpl$, where $\wc=Q B/m$ is the cyclotron frequency (magnetic field $B$), to values on the order of one. Of particular interest is the excitation spectrum perpendicular to the magnetic field, which features Bernstein modes in the weakly coupled domain and higher harmonics of the upper hybrid mode in strongly coupled systems. It is shown that Bernstein modes exist also in weakly magnetized plasmas at moderate coupling, when they are located below the upper hybrid mode. We further investigate the latter mode for a variety of coupling strengths in the $\Gamma\sim 1$ regime, and the transition to an unmagnetized plasma.

The work is organized as follows. In Sec.~\ref{sec:sim}, we introduce the simulation method and detail the numerical parameters. The results of the simulation are then presented in Sec.~\ref{sec:results}. We conclude with a brief summary and discussion in Sec.~\ref{sec:summary}.

\section{Simulation method}\label{sec:sim}
We briefly summarize the methodology of the simulations, which is the same as in our previous work~\cite{kaehlert2022prr}. We simulate the classical one-component plasma, i.e., $N$ identical particles with mass $m$ and charge $Q$ in a uniformly charged background, subject to an external magnetic field $\textbf{B}=B\hat e_z$. The particles are placed in a cubic box. Simulations are conducted with the LAMMPS code~\cite{LAMMPS,plimpton1995jcp}, thereby integrating the equations of motion with an extension of the velocity Verlet algorithm~\cite{spreiter1999}. We study the density autocorrelation function in Fourier space, i.e., the dynamic structure factor,
\begin{align}
    S(\textbf{k},\omega)=\frac{1}{2\pi}\int_{-\infty}^\infty F(\textbf{k} ,t) e^{i\omega t}dt,
\end{align}
where $F(\textbf{k},t)=\langle n_\textbf{k}(t)n^*_\textbf{k}(0)\rangle/N$. The DSF is computed from a Fourier transform of the microscopic particle density, $n_\textbf{k}(t)=\sum_{i=1}^N e^{-i\textbf{k}\cdot \textbf{r}_i(t)}$, where $\textbf{r}_i(t)$ ($i=1\dots N$) are the particle positions~\cite{kaehlert2019, kaehlert2022prr}.

As discussed above, the magnetization will be given as $\beta=\wc/\wpl$. The number of particles is set to either $N=80,000$ or $N=10,000$. In particular, in order to get access to (perpendicular) wave numbers $k_\perp$ that are small compared to the inverse Larmor radius, $\rL^{-1}=\wc/\vth$ [thermal velocity $\vth=\sqrt{\kB T/m}$], which is an important length scale for waves in the weakly coupled domain, simulations with high particle numbers are required as the minimum wave number, $k_\text{min}=2\pi/L$, is dictated by the size of the simulation cell $L$, with $L/a=(4\pi\,N/3)^{1/3}$ (for cubic boxes). This is particularly important for small $\beta$, as can be seen from $k_{\text{min}} r_\text{L}=k_\text{min}a \cdot (\rL/a)=2\pi\times[3/(4\pi\, N)]^{1/3}\times (\sqrt{3\Gamma}\beta)^{-1}$, which should satisfy $k_{\text{min}} r_\text{L}< 1$. The time step varies from $\Delta t\, \wpl=0.0015$ to $\Delta t \,\wpl=0.01$, depending on the coupling strength. We conduct multiple new simulations for weak magnetic fields but also analyze data produced in the simulation runs of Ref.~\cite{kaehlert2022prr}.

\section{Results}\label{sec:results}
Since the effect of the magnetic field on the DSF for wave vectors parallel to the external field is typically weak, in particular for $\beta<1$, we focus on the DSF for wave vectors $\textbf{k}$ perpendicular to $\textbf{B}$ in the following.

\begin{figure}
\centerline{\includegraphics{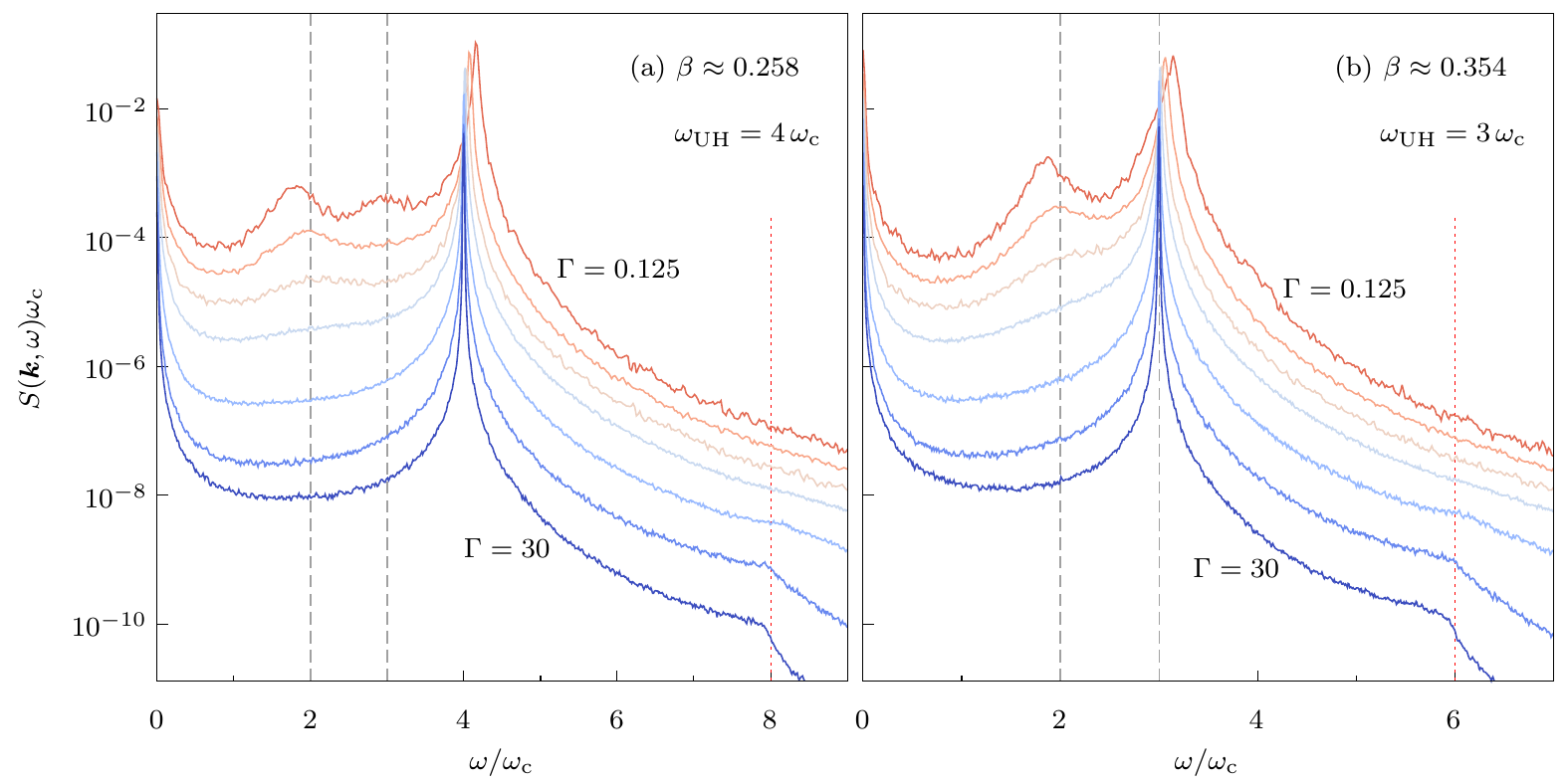}}
\caption{DSF for $\textbf{k}\perp \textbf{B}$ and coupling parameters $\Gamma\in\{0.125, 0.25, 0.5, 1, 3, 10, 30\}$ (from top to bottom). The perpendicular wave number is $k_\perp a=0.0905$. The magnetization is indicated in the figure. The vertical lines indicate harmonics of the cyclotron frequency (long dashed, grey) and the upper hybrid frequency (short dashed, red).\label{fig:bernsteinlowg}}
\end{figure}
Figure~\ref{fig:bernsteinlowg} depicts the transition of the DSF at a small wave number from weak to strong coupling in weakly magnetized plasmas, thereby complementing results in Fig.~3 of Ref.~\cite{kaehlert2022prr} with much stronger magnetization. According to the RPA (random phase approximation) description of collective modes in magnetized plasmas~\cite{bernstein1958pr,bellan2008book}, Bernstein modes below the upper hybrid frequency, $\omega_\text{UH}=\sqrt{\wpl^2+\wc^2}$, originate at $n\,\omega_\text{c}$ in the small wave number limit, with $n\ge 2$. This is clearly seen in Fig.~\ref{fig:bernsteinlowg}(a) for $\Gamma=0.125$, where two maxima are found below the upper hybrid frequency. The lower peak is located slightly below $2\,\omega_\text{c}$, which could be caused by thermal effects due to a finite $k_\perp r_\text{L}\approx 0.57$. The shift as well as the peak intensity become weaker and eventually disappear as the coupling strength increases. The DSF for plasmas with very strong coupling, $\Gamma\gtrsim 3$, exhibits a rapid decay for frequency above $2\,\omega_\text{UH}$, as observed previously~\cite{kaehlert2022prr}. Very similar behavior can be seen in Fig.~\ref{fig:bernsteinlowg}(b) with a slightly larger magnetization. Here, only one Bernstein mode exists below the upper hybrid frequency at weak coupling.

The discussion above shows that the observations made in Ref.~\cite{kaehlert2022prr} remain valid in weakly magnetized plasmas, i.e., the Bernstein modes that exist in weakly coupled systems disappear upon increase of $\Gamma$. When the plasma approaches the strongly coupled regime, the DSF decays rapidly for $\omega\gtrsim 2\omega_\text{UH}$, but, in the present simulations, the DSF does not (yet) display clear peaks around the harmonics of $\omega_\text{UH}$ due to the weak magnetization. The harmonics are significantly more pronounced when $\beta\gtrsim 1$ and $\Gamma\gg 1$~\cite{kaehlert2022prr}.

\begin{figure}
\centerline{\includegraphics{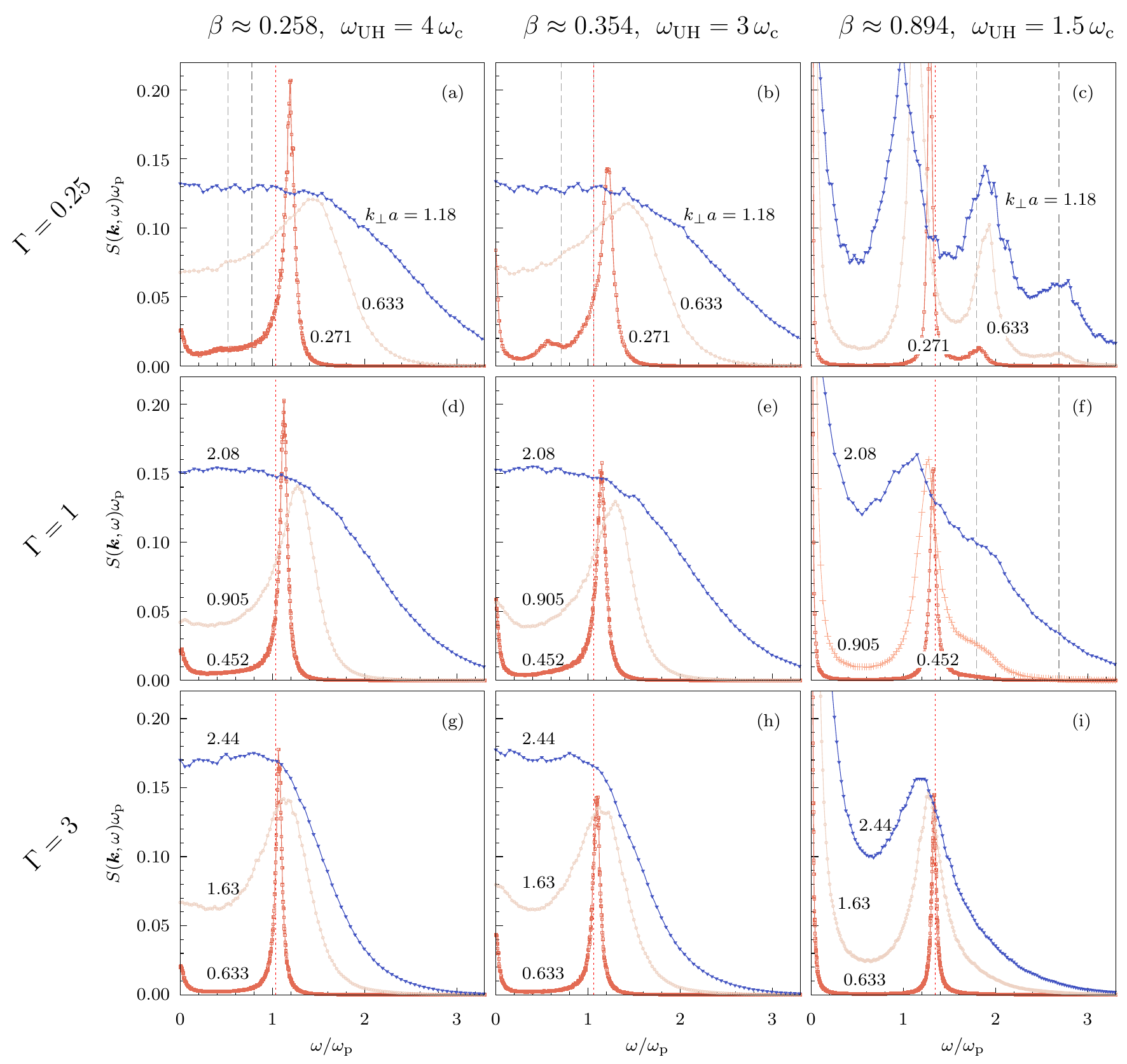}}
\caption{DSF for $\textbf{k}\perp \textbf{B}$ and coupling parameters $\Gamma\in\{0.25, 1, 3\}$ (top row to bottom row) and $\beta\in\{0.258,0.354,0.894\}$ (left column to right column). The perpendicular wave numbers $k_\perp a$ are indicated in the figure. The vertical lines show harmonics of the cyclotron frequency (long dashed, grey) and the upper hybrid frequency (short dashed, red).\label{fig:overview}}
\end{figure}
We now consider a larger range of wave numbers in Fig.~\ref{fig:overview}, for various coupling and magnetization strengths. Consider first the top row with $\Gamma=0.25$. For $\beta\approx 0.258$ [Fig.~\ref{fig:overview}(a)], the DSF has a clear peak at small $k_\perp$, which broadens and shifts to higher frequencies as the wave number increases. A weak remnant of the lowest Bernstein mode is visible at $k_\perp a=0.271$ near $2\,\omega_\text{c}$, see the dashed vertical line. At the largest wave number, $k_\perp a=1.18$, the main peak has disappeared, and the DSF decays monotonically, with a broad plateau region for $\omega\lesssim 1.4\,\wpl$. The lowest Bernstein mode becomes somewhat more pronounced as the magnetization is increased to $\beta\approx 0.354$ [Fig.~\ref{fig:overview}(b)]. At the largest magnetization [$\beta\approx 0.894$, Fig.~\ref{fig:overview}(c)], the Bernstein modes are located above the upper hybrid frequency. In this case, they are clearly visible in the DSF. In particular, for the largest wave number, the DSF no longer decays monotonically but is modulated by peaks around the harmonics of the cyclotron frequency. The upper hybrid peak shifts to lower frequencies with increasing wave number. As the coupling parameter is raised to $\Gamma=1$ (middle row), any obvious signature of the Bernstein modes is lost for $\beta\approx 0.258$ and $\beta\approx 0.354$. Only for $\beta\approx 0.894$, they remain detectable in the spectrum. At $\Gamma=3$ (bottom row), the upper hybrid mode is the only visible excitation in the spectrum, apart from a zero frequency peak, which generally becomes more dominant at larger $\beta$, see also $\Gamma=1$ and $\Gamma=0.25$.

\begin{figure}
\centerline{\includegraphics{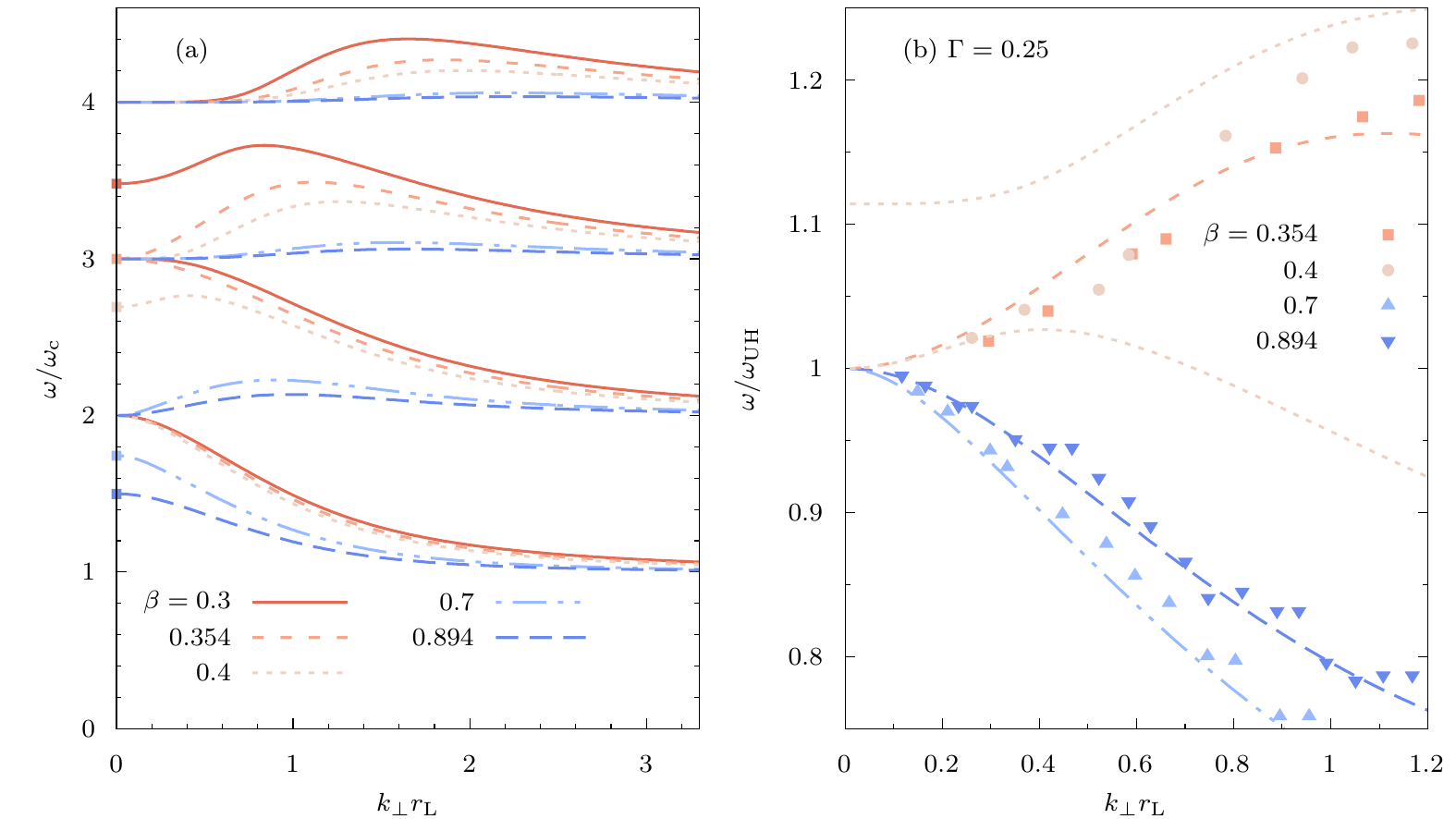}}
\caption{(a) RPA dispersion relation for the magnetized OCP for various levels of magnetization, as indicated in the figure. (b) Comparison of the upper hybrid dispersion from the RPA (lines) with the main peak position from the DSF (symbols). Note the different scaling of the vertical axis in (a) and (b).\label{fig:dispersion}}
\end{figure}
Considering the position of the main peak with respect to the upper hybrid frequency, the results in Fig.~\ref{fig:overview} suggest that the dispersion of the upper hybrid mode can change from positive at low $\beta$ to negative at high $\beta$, similar to the transition in the unmagnetized OCP upon increase of $\Gamma$~\cite{mithen2012aip,korolov2015cpp}. Before we discuss the simulation results in more detail, we first recall the modification of the dispersion relation in the random-phase-approximation (RPA)~\cite{bellan2008book}, shown in Fig.~\ref{fig:dispersion}(a). It is obtained from the condition $\epsilon_\text{RPA}(\textbf{k},\omega)=0$, where
\begin{equation}\label{eqn:RPA}
\epsilon_\text{RPA}(\textbf{k},\omega)=1+\frac{1}{k^2\lambda^2}\left[ 1+\sum_{n=-\infty}^\infty   \frac{\omega}{\omega-n\,\omega_\text{c}} I_n(\eta)e^{-\eta}\, \zeta_n \,Z(\zeta_n)\right]
\end{equation}
is the RPA dielectric function~\cite{ichimaru1973book,alexandrov2013}. Here, $\lambda=\vth/\wpl$ is the usual Debye length, $\eta=k_\perp^2 \rL^2$, $\zeta_n=(\omega-n\,\wc)/(\sqrt{2}|k_\parallel|v_\text{th})$, and $I_n(\eta)$ [$Z(\zeta_n)$] denotes the modified Bessel [plasma dispersion] function. For perpendicular wave propagation, the dispersion relation is to be determined from the $k_\parallel\to 0$ limit of Eq.~\eqref{eqn:RPA}~\cite{bellan2008book}, resulting in
\begin{equation}\label{eqn:dispersion}
1=2\sum_{n=1}^\infty \frac{I_n(\eta)e^{-\eta}}{\eta}\frac{n^2 \wpl^2}{\omega^2-n^2 \wc^2}.
\end{equation}
We are primarily interested in the dispersion of the mode that starts at the upper hybrid frequency. The location of the latter is indicated by the squares in Fig.~\ref{fig:dispersion}(a). For $\beta<1/\sqrt{3}\approx 0.577$ ($\beta>1/\sqrt{3}$), the upper hybrid frequency lies above (below) the first Bernstein mode, which starts at $2\,\wc$. From the examples shown, one observes that the dispersion of the upper hybrid mode is positive in the former case and negative in the latter. In fact, expanding Eq.~\eqref{eqn:dispersion} in powers of $\eta=k_\perp^2 \rL^2$, one finds $\omega^2(k_\perp)\approx \wuh^2 +3\, \wc^2\, k_\perp^2 \rL^2 /(1-3\beta^2)$, where the $\sim k_\perp^2$ term becomes singular at $\beta=1/\sqrt{3}$ and changes sign. In case the upper hybrid frequency exactly coincides with one of the cyclotron harmonics, there are two modes originating from the same frequency at $k_\perp=0$. As discussed above, for $\wuh=2\wc$ [$\beta\equiv 1/\sqrt{3}$], the previous expression diverges and one finds, instead, two modes with a linear dependence on $k_\perp$, namely $\omega^2(k_\perp)\approx (2\wc)^2 \pm \wpl^2 k_\perp \rL$, see Ref.~\cite{kaehlert2022prr} for a comparison with simulations. For $\wuh=3\wc$, the result is $\omega^2(k_\perp)=(3\wc)^2 + \wpl^2\left(\frac{3}{10}\pm \frac{\sqrt{186}}{20}\right)k_\perp^2 \rL^2$. For all other cases, $\wuh=n\wc$, with $n\ge 4$, the general expression for the upper hybrid dispersion remains valid, and the frequency of the second mode decreases $\sim (k_\perp \rL)^{2n-2}$.

Figure~\ref{fig:dispersion}(b) shows a comparison of the RPA dispersion and the peak position from the DSF for $\Gamma=0.25$. For $\beta=0.7$ and $\beta=0.894$, there is good agreement between theory and simulation, as observed previously in Ref.~\cite{kaehlert2022prr} for $\beta=1/\sqrt{3}$. However, larger deviations occur for $\beta=0.354$ and, in particular, for $\beta=0.4$. In the latter case, the peak position from the simulations increases monotonically with $k_\perp \rL$, with a leap around $k_\perp \rL \approx 0.75$. In contrast, the frequency of the upper hybrid mode from the RPA has a maximum at $k_\perp \rL \approx 0.4$ and generally lies well below the peak of the DSF, except for the smallest wave numbers. Interestingly, the RPA Bernstein mode above the upper hybrid mode is in reasonable agreement with the simulations for $k_\perp \rL \gtrsim 0.75$. Even though $\beta=0.354$ is only marginally smaller than $\beta=0.4$, the RPA spectrum is quite different [shown is the upper of the two modes that start at $\wuh=3\,\wc$] and agrees much better with the simulation data, which, on the contrary, are very similar to those for $\beta=0.4$. These observations and the small separation of the RPA modes in frequency space for $\beta=0.4$ suggest that both the upper hybrid mode and the nearby Bernstein mode may contribute significantly to the DSF. We keep in mind, however, that the coupling is already outside the regime where the RPA is expected to apply. We also reiterate that the peaks in the DSF do not necessarily correspond precisely to the frequencies of the collective modes~\cite{kryuchkov2019,hamann2020prb}.

\begin{figure}
\centerline{\includegraphics{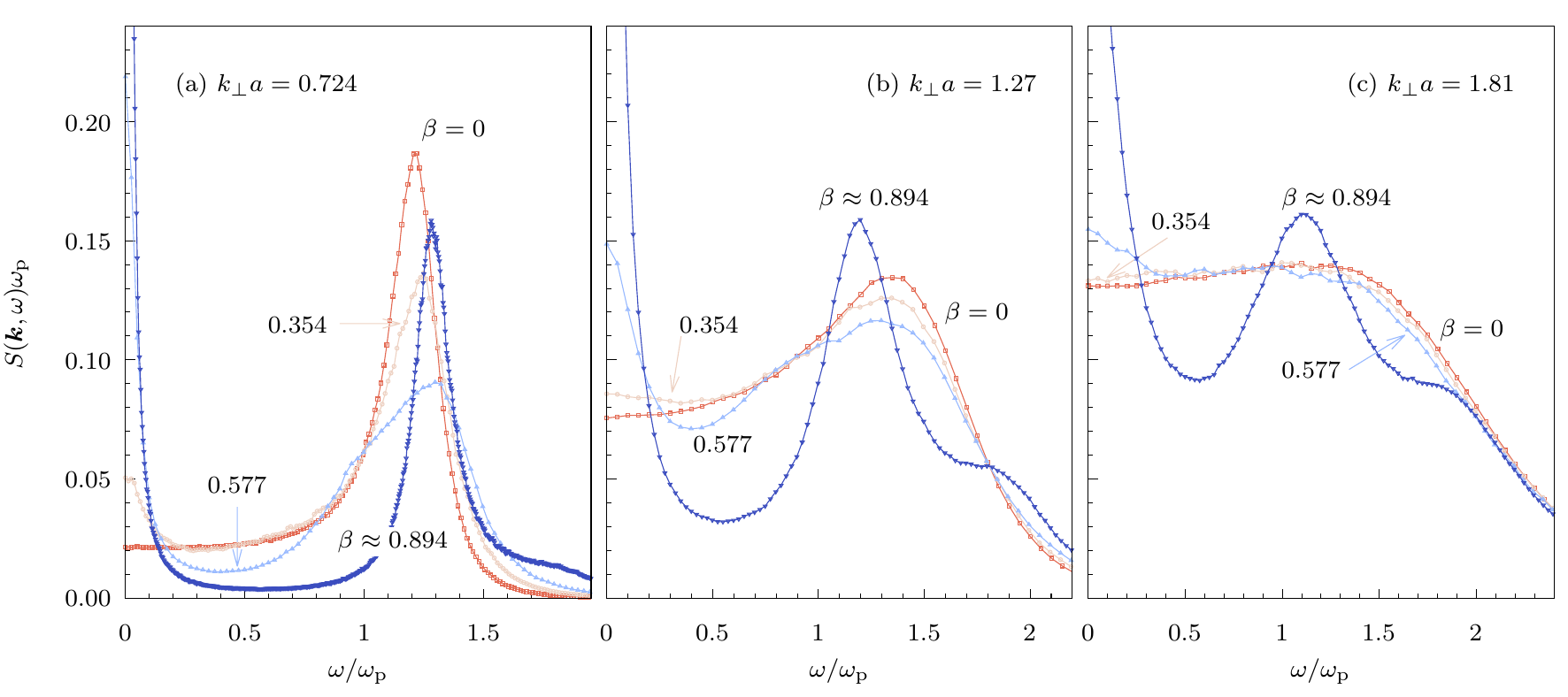}}
\caption{DSF perpendicular to the magnetic field for $\Gamma=1$ and various magnetizations $\beta$. Also shown is the unmagnetized limit with $\beta=0$. The wave number $k_\perp a$ is indicated in the figure.\label{fig:diffb}}
\end{figure}
Finally, the evolution of the DSF upon magnetizing the plasma is studied in Fig.~\ref{fig:diffb} for $\Gamma=1$ and three different wave numbers. At $k_\perp a=0.724$ [Fig.~\ref{fig:diffb}(a)], the plasmon (upper hybrid) peak for $\beta=0$ ($\beta>0$) initially becomes broader as $\beta$ increases while, at the same time, the peak height decreases. At larger magnetization, the trend is reversed. The peak becomes sharper, and the peak intensity grows again ($\beta=0.894$).  Since the normalization of the DSF is independent of $\beta$, $\int_{-\infty}^\infty S(\textbf{k},\omega;\beta)d\omega=S(k)$, where $S(k)$ is the static structure factor, the spectral weight contained in the plasmon merely becomes redistributed---the total weight remains constant. In addition to the plasmon/upper hybrid feature, a very sharp zero-frequency peak develops, even for relatively weak magnetization strengths, which is absent in the $\beta=0$ spectrum. For $k_\perp a=1.27$ [Fig.~\ref{fig:diffb}(b)] and $\beta=0$, the plasmon peak is already much broader, which is well known from simulations of the unmagnetized OCP~\cite{mithen2012pre}. As the plasma becomes magnetized, the peak shifts to lower frequencies. At $\beta\approx 0.577$, a strong zero-frequency peak has emerged. Increasing $\beta$ further to $\beta\approx 0.894$ leads to the formation of a well-pronounced upper hybrid peak. In case of the largest wave number, $k_\perp a=1.81$ [Fig.~\ref{fig:diffb}(c)], the plasmon peak practically vanishes for $\beta=0$. Similar to Fig.~\ref{fig:diffb}(b), increasing the magnetization leads to the formation of a well-pronounced peak in the DSF, both at zero and finite frequency. Comparing the effect of the magnetic field on the DSF for different wave numbers, one finds that the strongest modifications, compared to the zero magnetic field case, are found at small wave numbers. We note that the peak position of the DSF is above (below) the upper hybrid frequency for small (large) magnetization, as in Fig.~\ref{fig:dispersion}(b). For a related discussion of the harmonics in the magnetized and unmagnetized OCP, see also Fig.~4 in Ref.~\cite{kaehlert2022prr}.

\section{Conclusions}\label{sec:summary}
In summary, we have studied the DSF of the weakly magnetized OCP for wave vectors perpendicular to the magnetic field for a variety of coupling strengths and wave numbers. At small wave numbers, the DSF shows signatures of Bernstein modes even for weak magnetization, $0.25\lesssim \beta \lesssim 1$, provided the coupling is sufficiently low, $\Gamma\lesssim 0.5$. Bernstein modes disappear as soon as $\Gamma$ grows beyond $\Gamma\gtrsim 1-3$. In this strongly coupled regime, the DSF decays rapidly for $\omega>2\wuh$, but the magnetization is too low to observe clear higher harmonics of the upper hybrid mode as in Ref.~\cite{kaehlert2022prr}, where larger field strengths were considered.

The main peak in the DSF broadens as the wave number increases, similar to the plasmon peak for $\beta=0$. For large wave numbers, the DSF decays monotonically in the case of weak magnetization. For stronger magnetization, Bernstein modes can be observed that modulate the decay to high frequencies. In addition, a strong zero-frequency peak is observed, which appears already for relatively mild magnetization. At finite frequencies, the dispersion of the main peak was studied in some detail for $\Gamma=0.25$. It changes from positive to negative upon increase of $\beta$. Since it is known that the dispersion of the plasmon becomes negative for $\Gamma\gtrsim 10$ already in the unmagnetized OCP~\cite{mithen2012aip, korolov2015cpp}, this effect could vanish in the strongly coupled regime. Here, a theory for the dispersion must include correlations~\cite{khrapak2016pop}. The comparison between the RPA dispersion and the peak position from the simulations shows that, under certain conditions, the main peak in the DSF could be influenced by the interplay of the upper hybrid mode and a nearby Bernstein mode. This, however, requires further investigation.

The study of the DSF at a fixed wave number and intermediate coupling ($\Gamma=1$) shows that the plasmon/upper hybrid peak shifts and changes its shape upon increase of the magnetization. At small $k_\perp$, the peak initially broadens and its intensity decreases after the trend is reversed, and the peak sharpens again. At larger $k_\perp$, the plasmon peak for $\beta=0$ is much less pronounced. Increasing the magnetization, however, leads to the formation of well-developed upper hybrid features. In addition, a strong zero-frequency mode emerges at all considered wave numbers.

\section*{Acknowledgments}
The simulations were performed at the
Norddeutscher Verbund für Hoch- und Höchstleistungsrechnen (HLRN) under Grant No. shp00026.

\providecommand{\url}[1]{\texttt{#1}}
\providecommand{\urlprefix}{}
\providecommand{\foreignlanguage}[2]{#2}
\providecommand{\Capitalize}[1]{\uppercase{#1}}
\providecommand{\capitalize}[1]{\expandafter\Capitalize#1}
\providecommand{\bibliographycite}[1]{\cite{#1}}
\providecommand{\bbland}{and}
\providecommand{\bblchap}{chap.}
\providecommand{\bblchapter}{chapter}
\providecommand{\bbletal}{et~al.}
\providecommand{\bbleditors}{editors}
\providecommand{\bbleds}{eds: }
\providecommand{\bbleditor}{editor}
\providecommand{\bbled}{ed.}
\providecommand{\bbledition}{edition}
\providecommand{\bbledn}{ed.}
\providecommand{\bbleidp}{page}
\providecommand{\bbleidpp}{pages}
\providecommand{\bblerratum}{erratum}
\providecommand{\bblin}{in}
\providecommand{\bblmthesis}{Master's thesis}
\providecommand{\bblno}{no.}
\providecommand{\bblnumber}{number}
\providecommand{\bblof}{of}
\providecommand{\bblpage}{page}
\providecommand{\bblpages}{pages}
\providecommand{\bblp}{p}
\providecommand{\bblphdthesis}{Ph.D. thesis}
\providecommand{\bblpp}{pp}
\providecommand{\bbltechrep}{}
\providecommand{\bbltechreport}{Technical Report}
\providecommand{\bblvolume}{volume}
\providecommand{\bblvol}{Vol.}
\providecommand{\bbljan}{January}
\providecommand{\bblfeb}{February}
\providecommand{\bblmar}{March}
\providecommand{\bblapr}{April}
\providecommand{\bblmay}{May}
\providecommand{\bbljun}{June}
\providecommand{\bbljul}{July}
\providecommand{\bblaug}{August}
\providecommand{\bblsep}{September}
\providecommand{\bbloct}{October}
\providecommand{\bblnov}{November}
\providecommand{\bbldec}{December}
\providecommand{\bblfirst}{First}
\providecommand{\bblfirsto}{1st}
\providecommand{\bblsecond}{Second}
\providecommand{\bblsecondo}{2nd}
\providecommand{\bblthird}{Third}
\providecommand{\bblthirdo}{3rd}
\providecommand{\bblfourth}{Fourth}
\providecommand{\bblfourtho}{4th}
\providecommand{\bblfifth}{Fifth}
\providecommand{\bblfiftho}{5th}
\providecommand{\bblst}{st}
\providecommand{\bblnd}{nd}
\providecommand{\bblrd}{rd}
\providecommand{\bblth}{th}





\end{document}